\documentclass{JHEP3}

\usepackage{graphicx}
\usepackage{verbatim}
\usepackage{epsf}
\usepackage{latexsym}
\usepackage{amsmath,amsfonts,amssymb,amsthm}

\newcommand{\bbox}{\lower.2ex\hbox{$\Box$}}

\newcommand{\beq}{\begin{equation}}
\newcommand{\eeq}{\end{equation}}
\newcommand{\bea}{\begin{eqnarray}}
\newcommand{\eea}{\end{eqnarray}}
\newcommand{\ena}{\end{eqnarray}}

\newcommand {\non}{\nonumber}

\newcommand{\Tr}{{\rm Tr}}

\renewcommand{\[}{\left[}
\renewcommand{\]}{\right]}

\newcommand{\be}{\begin{equation}}
\newcommand{\ee}{\end{equation}}

\preprint{UCSD/PTH 10-12}

\title{\begin{center} 
Dynamical SUSY breaking  and the $\beta$-deformation
\end{center}}
\author{
Antonio Amariti$^{1,a}$
\\~
\\
$^1$Department of Physics, University of California\\
San Diego La Jolla, CA 92093-0354, USA
\\
 ~~\\
$^a$\email{amariti@physics.ucsd.edu} 
}

\abstract{
We study supersymmetry breaking metastable vacua
arising from beta deformed quiver gauge theories.
The relation between the bounds on metastability 
and the deformation are discussed. 
Metastable supersymmetry breaking vacua
are found in the IR of beta deformed cascading quivers
with vector-like field content.
Furthermore the limiting case of massive $N_f=N_c$
SQCD appears in the IR of gauge theories with chiral-like field content.
We comment on the field theory origin of the deformation and on
possible applications in AdS/CFT.
}
\begin{document}

\section{Introduction}

Dynamical supersymmetry breaking (DSB) \cite{Witten:1981nf,Affleck:1983mk,Affleck:1984xz}
 is a promising mechanism for the model building of testable models of physics 
 beyond the SM.
 The main difficulty in model building is that DSB usually appears at
strong coupling and the calculation of the spectrum is
very involved if not impossible \cite{Intriligator:1996pu}.
Nevertheless \cite{Intriligator:2006dd} showed that the analysis of the strongly coupled
phase can be avoided in some cases, 
by applying $\mathcal{N}=1$ UV/IR duality \cite{Seiberg:1994pq}. 
Indeed the low energy dual phase of a strongly coupled \emph{electric} theory is 
weakly coupled in the IR, and pertubatively accessible. By following the flow 
of  massive SQCD in the free \emph{magnetic} window in \cite{Intriligator:2006dd} 
it was shown that supersymmetry is dynamically broken in metastable vacua in the IR.
These vacua are stable and long living if the masses of the electric theory
are small enough. 
This fine tuning of the quark masses poses a problem of naturalness of
DSB in SQCD.

After \cite{Intriligator:2006dd} there have been some attempts to
find out a dynamical origin of this mass term, such to explain its smallness.
For example it is possible to make the mass term naturally small by retrofitting, i.e. by generating the SUSY breaking scale from the strong dynamics of some other gauge groups \cite{
Dine:2006gm,Aharony:2006my,Brummer:2007ns}.
Another related possibility is the suppression of the quark masses caused by stringy
instanton contributions \cite{Argurio:2007qk,Aharony:2007db}.
Most recently  in \cite{Brummer:2010zx} a different possibility of
supersymmetry breaking without scales has been proposed.
Basically one may obtain the ISS
model by starting with a theory without a relevant mass scale and then by acting with multiple
Seiberg dualities.
The action of the dualities reduces the number of elementary operators in some
superpotential deformations. Even if there are not  relevant massive deformations in the
UV, the mass terms are generated dynamically in the IR after multiple dualities.
Usually  these masses are proportional to the strong gauge coupling scales
and they do not lead to ISS vacua,
because they are too large and they are integrated out in the IR.
Anyway in \cite{Brummer:2010zx} it was argued that by imposing some constraint on the UV 
marginal  superpotential couplings, then the IR mass terms can be  small enough to ensure the existence of ISS vacua.

In this paper we  generalize the idea of metastable vacua without UV scales 
to quiver gauge theories arising from D3 branes at toric $CY_3$ singularity.
Quiver gauge theories have been a large laboratory for the investigation
of metastable supersymmetry breaking in the last years \cite{Franco:2006es}.

Here, 
instead of tuning the UV couplings by hand, we restrict to a
submanifold of the conformal manifold $\mathcal{M}_c$
in which the correct hierarchy is automatically provided.
This submanifold 
is found along the directions of $\mathcal{M}_c$ corresponding 
to the $\beta$-deformation.
Indeed by $\beta$-deforming the model and by changing the ranks of some of the gauge groups (consistently with the anomaly equations)
 we generate a RG flow that leads in the IR to ISS vacua,  without turning on any UV
 massive  deformation.
This mechanism potentially generates the ISS mass term even in chiral 
gauge theories,  where otherwise it is impossible to switch on relevant massive
deformations by hand. 

Another interesting aspect of beta deformed supersymmetry breaking is that 
in principle it is possible to study the gravity dual of these metastable vacua.
Indeed the supergravity solutions dual to these quiver gauge theories are known,
and they are represented by D3 branes on 
AdS$_5\times$ Y$^5$ (where $Y^5$ is a Sasaki-Einstein manifold).
Moreover both the beta deformation and the breakdown of the conformal invariance
can be easily controlled in the dual gravity theory. Indeed
from the gravity side the beta deformation correspond to a TsT or STsTS$^{-1}$
\cite{Lunin:2005jy} 
transformation of the background, and the breakdown of the conformal invariance 
is associated to the addiction of fractional branes wrapped on the cycles of the geometry.

In section \ref{section1} we discuss the mechanism to generate ISS vacua in beta deformed toric
quiver gauge theories. In section \ref{section3} we study a vector-like example
coming from a $\mathbb{Z}_2$ orbifold of the conifold
in which the IR coincides with massive  SQCD in the free magnetic  region.
In section  \ref{section4} 
we study chiral models in which the IR dynamics is given by $N_f=N_c$
massive SQCD, the limiting case discussed in \cite{Intriligator:2006dd}. 
In section \ref{section6} we discuss the constraints imposed by field theory on the $\beta$ deformation and the implications for our results.
In section \ref{section7} we review the geometrical aspects of the deformation and 
we speculate on the connection with this mechanism and the AdS/CFT duality.
Then we conclude.

\section{The general setup}
\label{section1}

\begin{figure}
\centering
\includegraphics[width=8cm]{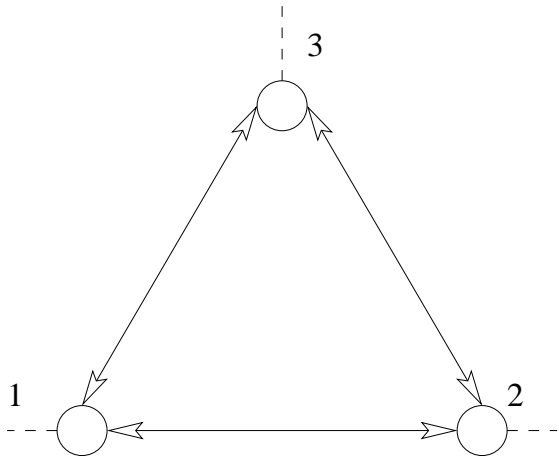}
\caption{Relevant part of the quiver that generate the ISS mechanism}
\label{fig1}
\end{figure}

In this section we discuss the mechanism to generate
no scale supersymmetry  breaking metastable models from D3 branes at singularities.
We start with a toric quiver gauge theory with a marginal superpotential.
In Figure \ref{fig1} we isolate the relevant part of the quiver necessary to our discussion.
This is the sector of a generic quiver in which the masses of the quarks are generated by
duality.
This sector in the IR has the same field content and symmetries of the ISS model and generates
the metastable vacua.
We suppose that the rest of the quiver does not spoil the result. For example there are no
pseudo-flat directions with negative quantum corrections and tree level 
tachyons. In the rest of the paper 
we study  many examples   where the validity of this assumption can be checked.
We suppose that this quiver has a superpotential
\begin{equation} \label{spot00}
W = h(X_{12} X_{23} X_{31} - X_{21} X_{13}X_{32}) 
\end{equation}
Toric quiver gauge theories usually admit  marginal deformations. We study one
of them, which is identical to the  superpotential, except that all the coupling are the same.
This corresponds to the $\beta$ deformation originally defined in $\mathcal{N}=4$ SYM \cite{Leigh:1995ep}.
As discussed in \cite{Benvenuti:2005wi} in toric quiver gauge theories this deformation
 is always marginal, because it commutes with the Cartan generators of the global symmetries. 
Usually toric quiver gauge theories have a global  $U(1)^3$, which corresponds 
to the isometries of the basis of the Sasaki Einstein manifold.
This symmetry corresponds to $U(1)_F^2 \times U(1)_R $ in the field theory, and it can be   enhanced by some non abelian factor (as in the $Y^{pq} $ case), or by some baryonic symmetry.
Anyway the beta deformation always preserves the $U(1)_F^2$ symmetry and for this reason it
is always exactly marginal
\footnote{Indeed as claimed in \cite{Green:2010da} indeed any marginal operator invariant under the global symmetries is exactly marginal}.
The superpotential (\ref{spot00}),  modified by the marginal $\beta$ deformation
\footnote{Observe that $h$ and $\lambda$ are complex numbers and there is not,
in general, a field redefinition such that they can both be considered real. It follows that the 
couplings $f_{\pm}$ have to be treat as complex too.}, becomes
\begin{equation}
W = (h+\lambda) X_{12} X_{23} X_{31} - (h-\lambda) X_{21} X_{13}X_{32} \equiv
f_+ X_{12} X_{23} X_{31} -f_- X_{21} X_{13}X_{32}
\end{equation}
Usually these theories admit a larger set of conformal deformations . In particular as shown in \cite{Benvenuti:2005wi,Green:2010da,Kol:2002zt,Kol:2010ub} it depends from the number of super-marginal operators coupled to
the conserved currents associated to the global symmetries.
In this paper we ignore the other deformations, but they can be useful to enforce other interesting properties as R symmetry breaking.
Then we choose an assignment of ranks (different number of fractional D5 branes
wrapped on 2 cycles collapsed at
the CY singularity), such that one of the groups has
$N_f=N_c$. For example we choose $N_1=N_2+N_3$. This groups is described at low energy by an effective theory of mesons and baryons, with the quantum modified moduli space.
The superpotential becomes
\be
W = f_{+} \Lambda_1 X_{23} M_{32} + f_{-} \Lambda_1 X_{23}M_{32}+
\Lambda \left(\det \left(
\begin{array}{cc} 
M_{22} & M_{23}\\
M_{32}& M_{33}
\end{array}
\right)
 -B_{23} B_{32} - \Lambda_1^{2 N_1} 
 \right) 
\ee
where $\Lambda$ is a Lagrange multiplier enforcing the quantum correction of the moduli space.
We choose a solution of this constraint in the baryonic branch, such that $B_{23} B_{32} =\Lambda_1^{2 N_1} $ which leaves the condition $\det M= 0$. 
If we look at the first two mass terms in the superpotential we notice that an appropriate choice of $h \simeq \lambda$ gives us one large mass term, which can be integrated out, while the 
other term remains small (namely $|f_+| \gg |f_-|$).
We are left with an $SU(N_2) \times SU(N_3)$ gauge theory with superpotential
\be
W = f_{-} \Lambda_1 M_{23} X_{32} 
\ee
If $N_{2} <N_{3} < 3/2 N_2$ we have the same field content of the ISS model
but  the theory has two gauge groups. If we suppose that $\Lambda_2 \ll \Lambda_3$ we can think the $SU(N_2)$
gauge group as a spectator and the dynamics of this group is negligible at this scale.
Seiberg duality on the $SU(N_3)$ group gives  the ISS model, which breaks supersymmetry in metastable vacua.
Basically this  shows that it is possible, by starting from some point in the conformal manifold
of a SCFT, to end up with a SUSY breaking gauge theory, just by
enforcing a cascading RG  flow with fractional branes. 
It is an example of  supersymmetry breaking without UV mass terms, 
in which the hierarchy required on the small
mass term in ISS  corresponds to an opportune choice of 
a submanifold of $\mathcal{M}_c$, where $h \simeq \lambda$.

In the rest of this section we generalize the mechanism explained for
the quiver in Figure \ref{fig1} to more generic models.
We start from the superpotential
\begin{equation}
W = h_{\[i_{1},\dot i_{n}\]}\mathcal{O}_{i_{1}\dots\, i_{n}} + \lambda _{\{i_{1},\dot i_{n}\}}\mathcal{O}_{i_{1}\dots\, i_{n}}
\end{equation}
where the parenthesis $\[,\]$ and $\{,\}$ represent the anti-symmetric toric superpotential and the symmetric $\beta$ deformation. The couplings become
$f_\pm=h\pm \lambda$ and we require $h \simeq \lambda$.
This mechanism generates the 
small couplings in the superpotential necessary for the ISS mechanism.
At this point of the discussion we can break the conformal invariance 
by changing the ranks of the gauge groups.
If the quiver has a vector like structure we can assign the ranks  without any constraint from the gauge anomalies, otherwise, if there is a chiral-like matter field structure, the constraint
from the gauge anomalies has to be imposed on the choice of the ranks.
Some of the groups become strongly coupled in the  IR. 
Naively we choose an assignation of ranks such that 
the number of these groups corresponds to the number of operators 
appearing in $\mathcal O$ lowered by two units \footnote{This is true if the fields in 
$\mathcal{O}$ are connected with all that groups. Otherwise there can
be other groups which can run to strong coupling too.}. 
We suppose that there is an anomaly free assignation of gauge group ranks
such that the groups running faster to strong coupling have
$N_f = N_c$. At low energy they vanish and they are replaced by 
a theory of mesons and baryons with a quantum
modified moduli space. On the baryonic branch we observe that
we are left with a low energy theory with an ISS sector and, by imposing the correct
hierarchy on the scales of the gauge groups, long living metastable vacua are obtained.
In the next sections we study this mechanism at works in some toric
quiver gauge theories.

 \section{The double conifold}
 \label{section3}
 \begin{figure}
\centering
\label{fig5}
\includegraphics[width=12cm]{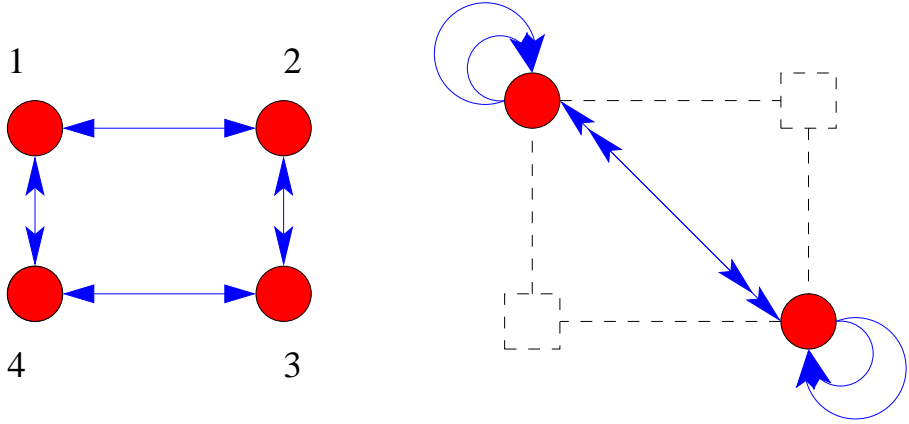}
\caption{The UV double conifold quiver and the IR version after nodes $1,3$ have
flown to strong coupling. 
The dashed lines are associated to the baryons, and naively the dashed squared can
be thought as the gauge groups which have disappeared in the IR.}
\end{figure}
The first example that we discuss 
is a $\mathbb{Z}_2$
 orbifold of the conifold, called double conifold or $L^{222}$ theory.
 The theory consist of four $SU(N)$ gauge groups, the bifundamental have a vector-like structure and the superpotential is
 \begin{equation} \label{spotL222}
 W/h  = X_{12}X_{23}X_{32}X_{21}-X_{23}X_{34}X_{43}X_{32}+X_{34}X_{41}X_{14}X_{43}-X_{41}X_{12}X_{21}X_{14}
 \end{equation}
 In Figure \ref{fig5} we give a representation of the quiver associated to the superpotential 
 (\ref{spotL222}).
 By adding the beta deformation the interactions don't change but the couplings in(\ref{spotL222}) 
 are modified as above, by defining $f_{\pm}=h\pm \lambda$.
This theory is a SCFT and the RG flow is generated by breaking the symmetry among the gauge groups. We choose an assignation of fractional branes such that there is a cascading RG
flow which can end with ISS vacua.
For example here we choose  
\begin{eqnarray}
 N_2  =  N_4  =  N_1+N_3
\end{eqnarray}
 The second and the fourth group are $N_f= N_c$ gauge theories, and they are described by the quantum modified moduli space.
On the baryonic branch we are left with a $SU(N_1) \times SU(N_3)$ theory with
 superpotential
 \begin{equation} \label{IRspot}
W= f_{-} ( \Lambda_2^2 M_{13}M_{31}  +
 \Lambda_4^2 N_{13}N_{31}   ) \!
 + f_{+}  \Lambda_2 \Lambda_4 \,(M_{11}N_{11}  +
 M_{33}N_{33}  )
  \end{equation} 
Notice that in this case the mass terms are proportional to $f \Lambda^2$.
Apparently they have mass dimension $2$ instead of $1$. This is because the couplings  $f$ 
are dimensionless in the UV. 
Anyway the fractional branes generate an RG flow 
towards the IR and these couplings acquire dimension $-1$. Indeed the gauge
 groups flow to strong coupling and they describe an effective theory of mesons and baryons.
The UV hierarchy $f_{-} \ll f_{+}$ between the UV dimensionless couplings is
preserved in the flow to the IR and  the masses proportional to $f_{+}$ in (\ref{IRspot}) are integrated out. Moreover,
 by imposing a hierarchy  among the strong coupling scales, $\Lambda_2 \ll \Lambda_4$,  we can integrate out the term $ N_{13}N_{31} $ too. We are left with a massive  $SU(N_1) \times SU(N_3)$ gauge theory.
There are no restrictions on the choice of the 
ranks $N_1$ and $N_3$ because of the vector like structure of the quiver.
Let's suppose that $N_1 < N_3 < 3/2 N_1$
and that $\Lambda_1 \ll \Lambda_3$. This theory becomes $SU(N_3)$ SQCD in the magnetic free window with massive quarks. If $f_{-} \Lambda_2^2 \ll \Lambda_3$ 
this low energy theory is the ISS model.
 
\section{Chiral models}
\label{section4}
The study of ISS vacua in chiral quiver gauge theory is usually problematic, because 
the mass terms cannot be generated in these theories.
Indeed the chiral structure of the interactions  is not compatible with 
relevant mass deformations.
Anyway chiral gauge theories are very common CFTs in high energy
field theory. They are ubiquitous even in the AdS/CFT duality, 
and it should be useful to find out a mechanism that generates weakly 
coupled low energy DSB starting from these UV field theories.

As the beta deformation generates SUSY breaking without UV masses,
it can be applied to the chiral theories.
 Indeed  in this section we show that the ISS model can follow by beta-deforming 
 chiral toric gauge theories.
We discuss two explicit examples, $\mathbb{F}_{0}$ and dP$_2$, and we see that in both cases we end up with $N_f=N_c$ massive SQCD.
This theory is conjectured in ISS to have metastable supersymmetry breaking state.
We expect that some more exotic constructions may give origin to 
the ISS model in the magnetic free window ($N_f>N_c+1$), 
but we leave this topic for future works.

\subsection{$\mathbb F_0^{(I)}$}
\begin{figure}[htbp]
\centering
\label{fig4}
\includegraphics[width=12cm]{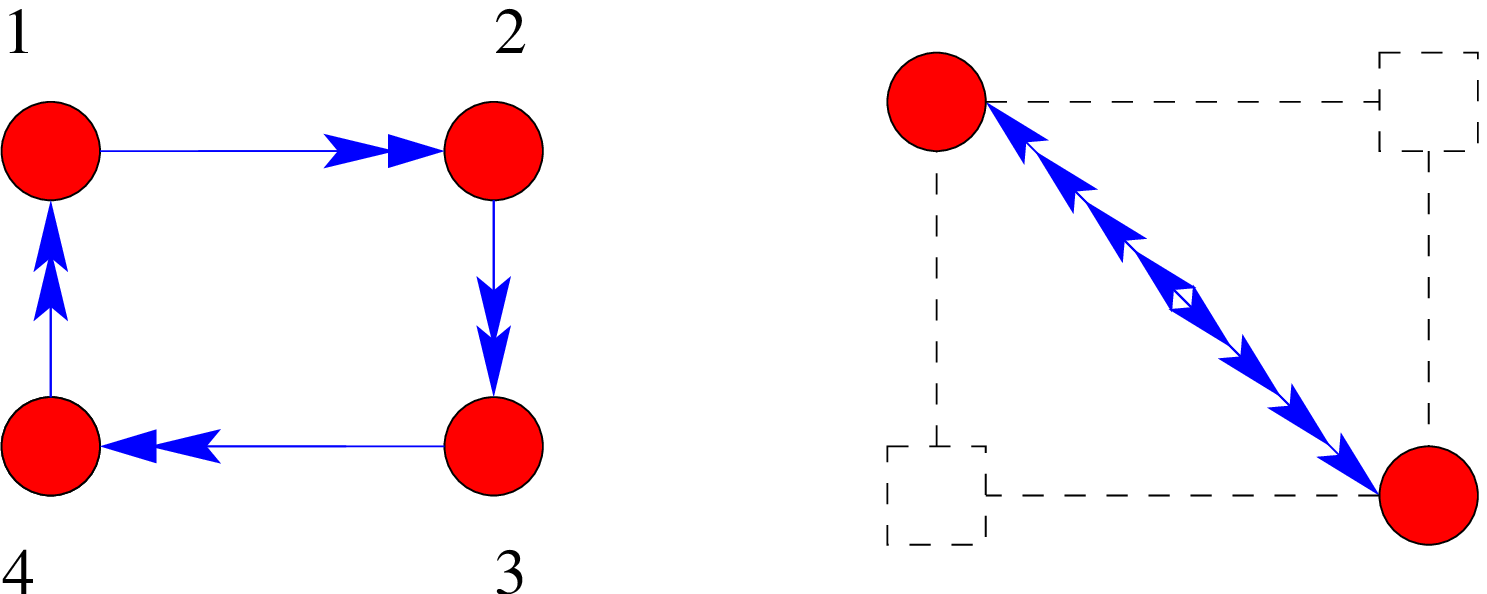}
\caption{The UV $\mathcal{F}_0^{(I)}$ quiver and the IR version after nodes $1,3$ have
flown to strong coupling. 
The dashed lines are associated to the baryons, and naively the dashed squares can
be thought as the gauge groups which have disappeared in the IR.}
\end{figure}
The first example that we study is the 
dual gauge theories of the complex cone over the Hirzebruch surface.
The models is usually referred as 
$\mathbb{F}_0^{(I)}$. This is a toric quiver gauge theory 
with chiral-like matter fields, $SU(N)^4$ gauge group and superpotential
\begin{equation}
W/h = \epsilon_{ij} \epsilon_{lk} X_{12}^{(i)} X_{23}^{(l)} X_{34}^{(j)} X_{41}^{(k)}
\end{equation}
In this case the only consistent assignation of fractional branes is $(M,K,M,K)$. We choose an assignation such the groups $N_2$ and $N_4$ have $N_f=N_c$. We are left with a 
$SU(N_1) \times SU(N_3)$ gauge theory with superpotential
\begin{equation}
W =   \Lambda_4 (\Lambda_2(f_{-} M_{13}^{(1)} M_{31}^{(1)} + f_{-} M_{13}^{(2)} M_{31}^{(2)}  )+ 
M_{13}^{(1)} M_{31}^{(2)} +M_{13}^{(2)} M_{31}^{(1)} ) 
\end{equation} 
Except from $M_{13}^{(1)} M_{31}^{(1)} $ the other fields are integrated out because their mass terms are very large. We are left with an $N_f=N_c$ gauge theory with a mass deformation, which in 
\cite{Intriligator:2006dd} is conjectured to break supersymmetry.

\subsection{dP$_2$}
\begin{figure}
\centering
\includegraphics[width=8cm]{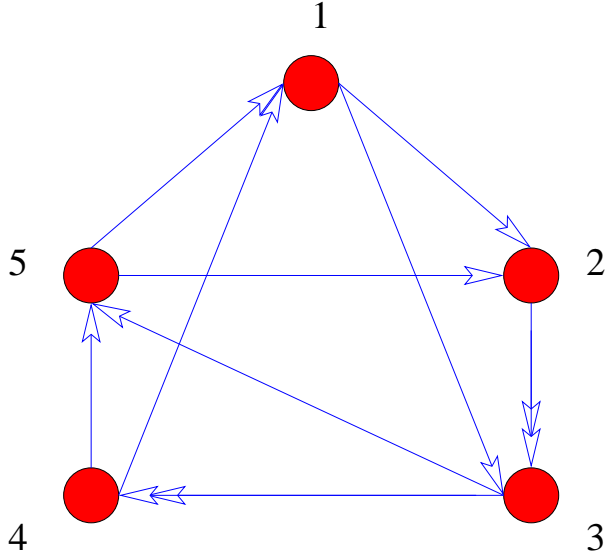}
\caption{The UV dP$_2$ quiver.}
\label{figdP2}
\end{figure}
Another chiral model with metastable supersymmetry breaking in the IR is the gauge theory
associated to the complex cone over the second del Pezzo surface, dP$_2$. This gauge theory has a chiral-like field content, $SU(N)^5$ gauge symmetry and it is described by the quiver in Figure 
\ref{figdP2}. The  superpotential is
\bea
W_{dP2} /h&=& X_{12}X_{23}X_{34}X_{45}X_{51} - X_{12} Y_{23} X_{35} X_{51} + 
X_{23}Y_{34}X_{45}X_{52} \non \\
&-& Y_{23} X_{35}X_{52}+ X_{13} Y_{34} X_{41} -
X_{13} X_{35} X_{51}
\eea
The beta deformation divides the couplings in 
$f_{+}$ and $f_{-}$. In particular here we associate $f_{-}$ to the coupling 
$X_{13} X_{35} X_{51}$.
We choose an assignation of fractional branes as 
$(N_1, N_2, N_3, N_4,N_5) = M (0,1,1,0,1)$. 
In such a way we reduce the theory to a $SU(N)^3$
gauge theory with superpotential $W= f_- \,( X_{13} X_{35} X_{51})$. 
By imposing a hierarchy among the scales $\Lambda_5 \gg \Lambda_3, \Lambda_1$
we first reduce to $SU(N_1) \times SU(N_3)$ with a mass term, while the group $N_5$ 
becomes an effective theory of mesons and baryons with a quantum modified moduli space. 
By assuming $ \Lambda_3 \gg  \Lambda_1$ we are left with an $SU(N_3)$ 
massive SQCD with $N_f=N_C$.

\section{Complex $\beta$ deformation}
\label{section6}
 
 In this section we study the field theory origin of the hierarchy among $f_{+}$ and 
 $f_{-}$.
 We start by writing the $\beta$ deformation with the notations of \cite{Leigh:1995ep}. In this case the
 superpotential for the $\mathcal{N}=4$ $\beta$-deformed field theory is usually
 written as
 \begin{eqnarray}
W =  h(e^{2 \pi i \beta } \Tr \Phi_1 \Phi_2 \Phi_3 -e^{-2 \pi i \beta } \Tr \Phi_1 \Phi_3\Phi_2) 
\end{eqnarray} 
The gauge coupling $g$ is real while $h$ is a complex coupling. The $\beta$ deformation
can be real or complex. 
For real values of the deformation parameter $\beta$ it was shown 
\cite{Leigh:1995ep} that there is a constraint 
among the three parameters,  
$\gamma(g,h,\beta)=0$. This constraint can be exactly solved in the large $N$ limit and it
implies $|h|=g$ with arbitrary real $\beta$.

In the case of complex beta the situation is more intricate. There is a large debate in the literature 
on the finiteness request and the conformal invariance of the complex beta deformation \cite{debate}.
Here we just require that the theory is conformal if 
the beta functions of the coupling constants are  vanishing and
we do not worry about the finiteness problem associated to the deformation. Moreover
we are actually interested in the breakdown of the conformal invariance, through the addition of the  
fractional branes. 
 Anyway, for complex beta the exact form of the $\gamma$ function is not known. In perturbation theory this problem has been studied in many papers 
 \cite{Freedman:2005cg,Mauri:2005pa,Khoze:2005nd}.
 For example at the leading order in $g^2\ll1$ in the planar limit
  it was found that   
  \be \label{rel}
  |h|^2 \cosh(2 \pi \text{Im} \beta) = g^2
  \ee
  We observe that our analysis requires complex $\beta$. Indeed if we consider a superpotential of the form
\be
W = h( e^{2 \pi i \beta} X_{ab} X_{bc} X_{ca} -e^{-2 \pi i \beta} X_{ba} X_{ac} X_{cb} )
\ee
a duality on  $N_a$ gives the superpotential 
\be \label{masse}
W = h \Lambda_a ( e^{2 \pi i \beta}  M_{cb} X_{bc} -e^{-2 \pi i \beta} M_{bc}  X_{cb} )
\ee
and the hierarchy among the two masses can be enforced only if Im($\beta$) 
is not too small.
By looking at (\ref{rel}) we observe that for small $g/|h|$  the value of 
Im($\beta$) is large enough to ensure a large hierarchy among the two masses
in (\ref{masse}), i.e. to impose $f_{-} \ll f_{+}$. 
We suppose that even at strong coupling (large $g$ and $|h|$) there is a similar relation among 
$g,h$ and Im($\beta$) such that we can still impose the constraint between $f_{+}$ and $f_{-}$

\section{ISS vacua and the gravity dual}
\label{section7}

In this section we discuss the existence of metastable supersymmetry breaking vacua from the perspective of the gauge/gravity duality.
Indeed the main ingredients that we used to generate these metastable states
in field theory are well understood from the dual gravity side.
For example the $\beta$ deformation can be implemented in every background with
an $U(1) \times U(1)$ non-R global symmetry, as showed in \cite{Lunin:2005jy}.
It consists of a TsT transformation,
two T dualities and a coordinate transformation on the metric.
The first T duality is done along a $U(1)$ direction and the shift is along an angle associated to the other $U(1)$.  This is the step at which the deformation parameter appears.
Then the last T duality is on the initial angle.
It is important to stress that since these transformation are performed on
the internal space, they are applicable even if the conformal symmetry is broken.
After beta-deforming the background  the dual toric geometry
still admits a complex structure deformation. 
When fractional D5 are added  the theory 
 has a cascading behavior and the beta deformation
has the same RG behavior of the original superpotential.
This suggests that the addition of the fractional branes doesn't affect the 
deformation parameters, and the UV hierarchy  among $f_{+}$ and $f_{-}$
is preserved during the cascading RG flow.

From the request of $U(1) \times U(1)$ symmetry it follows that  the beta deformations exists
for  every toric gauge theories, because they always admit a global $U(1)_F^2 \times U(1)_R$.
Differently from the case of AdS$_5 \times$ S$^5$, in the case of Sasaki-Einstein manifolds supersymmetry is
preserved if  the choice of the angles is done such that, after the shift, the Killing spinors remain invariant \cite{Minasian:2006hv,Butti:2007aq}.
Notice that the LM procedure works even in the case of complex $\beta$, where
the transformation is
slightly modified from TsT to STsTS$^{-1}$ where S denotes an $SL(2;R)$
trasformation.
In any case the deformation parameter has to be chosen such that $|\beta| \ll 1$.
This is in contrast with the requirement that we made in this paper, where the
imaginary part of $\beta$ has to be at least $\mathcal{O}(1)$.
Nevertheless, by following \cite{Chu:2006tp} we can allow larger values of $\beta$ by matching the periodicity property of $\beta$ in the dual gauge theory.  The deformation
is added in the dual string background more than in the supergravity approximation. In this case
the deformation arises from the string theory effective action, to all orders in $\alpha'$

The appearance of a complex beta deformation can be also related to the action of
the modular $SL(2,Z)$ groups. Indeed as argued in \cite{Intriligator:1999ff,
Dorey:2002pq,Dorey:2003pp}  the deformation parameter beta 
transform as a modular form under the modular action of the complexified gauge coupling.
This action of the S-duality group is strongly related to the requirement of a duals string background
\footnote{We thank the referee for focusing our attention to this connection between a complex beta deformation and
the action of the S-duality group.}.
It should be interesting to study the existence and the stability of the ISS vacua in a dual string 
background. We leave this analysis for future investigations.

\section{Conclusions}

In this paper we discussed the existence of  ISS vacua in quiver gauge theories without mass deformations.
We showed that these vacua arise by moving on some specific region of the conformal manifold
and then by relevantly deforming the theory with fractional branes.
They generate a cascading like behavior, which transforms some of  the superpotential 
deformations in mass terms. The specific region of the conformal manifold that we choose
doesn't allow integrate out all the massive fields generated by duality. 
The IR gauge theory becomes a SQCD-like model with massive quarks, and the ISS mechanism
can be recovered.

Many extensions are possible. 
First, to give some phenomenological relevance to this mechanism, it is necessary to break
$R$ symmetry. This breakdown can be implemented by other marginal deformation (indeed we are not 
spanning all the conformal manifold with the beta deformation). It is possible that they give some
higher power contribution to the superpotential, that can break R symmetry explicitly as in 
\cite{Amariti:2006vk}.

Breaking $R$ symmetry is  important if supersymmetry breaking has to be mediated to the MSSM
by a flavor blind mechanism, as gauge mediation. This mechanism is based on the gauging of the 
global flavor symmetry. In this case this symmetry is already gauged, and we already imposed 
some constraints on the strong coupling scales of the theory. It is possible that by adding some extra matter, charged under some symmetry of the supersymmetry breaking sector, some of the requirements that we did before have to be modified.
We leave this problem for future works.

Last, this mechanism seems promising for the study of ISS in AdS/CFT. Indeed, while adding
mass terms to the field theory can be quite complicate in supergravity duals, the addition of the  beta deformation is straightforward in the gravity dual description of toric quiver gauge theories.  As we discussed above, the requirement of large imaginary part for the deformation is the main obstruction for this analysis, because a string background and not its supergravity effective action is required.

\section*{Acknowledgments}

It is a great pleasure to thank Ken Intriligator  for comments on the draft. 
We also thank Alberto Mariotti for many discussions and 
CarloAlberto Ratti for useful clarifications. A.A. is
supported by UCSD grant DOE-FG03-97ER40546;.

\end{document}